\documentclass{PoS}

\usepackage{amsmath}

\newcommand{\dbar}{\mathchar'26\mkern-9mu d}
\renewcommand{\eqref}[1]{Eq.~(\ref{#1})}
\newcommand{\eqsref}[1]{Eqs.~(\ref{#1})}

\title{Including four-gluon interactions into Landau and 
maximally-Abelian gauge Dyson-Schwinger studies}

\ShortTitle{Including 4-gluon interactions into 
 Dyson-Schwinger studies}

\author{\speaker{Valentin Mader}\\
        U. Graz\\
        E-mail: \email{valentin.mader@uni-graz.at}}

\author{Reinhard Alkofer\\
       U. Graz\\
       E-mail: \email{reinhard.alkofer@uni-graz.at}}

\abstract{In Dyson-Schwinger studies of the Yang-Mills propagators
the four-gluon interaction has been usually neglected due to the related 
technical difficulties with the associated two-loop terms and especially 
their renormalization. A possible scenario to renormalize these fully-dressed 
two-loop terms is presented. Preliminary results for the Landau gauge gluon 
propagator are shown. Implications for the gluon propagators in maximally 
Abelian gauge are discussed.}

\FullConference{Xth Quark Confinement and the Hadron Spectrum\\
         October 8-12, 2012\\
         TUM Campus Garching, Munich, Germany}

\begin{document}

  \section{Motivation}

      Dyson-Schwinger equations (DSEs) provide a non-perturbative tool for studying
quantum field theories. Being an infinite set of coupled equations for the
$n$-point functions of the theory,  they contain in principle all information about
the observables of the underlying theory. In addition, since they are derived from
the renormalized action, they are fully renormalized equations and, given a certain
regularization and renormalization scheme, all divergences are absorbed in
appropriately determined renormalization constants.

      In practical calculations, however, truncations of the full infinite system
to a finite subset of equations have to be performed. In general these truncations
will interfere with the renormalization of DSEs and divergences will reappear. In
addition an inappropriate choice of a regulator might also introduce divergences,
as {\it e.g.}, a hard cut-off in numerical calculations. Thus even though the full
infinite set of DSEs might be fully renormalized, in practical calculations
divergences have to be properly subtracted. Furthermore, perturbative 
renormalization is  not sufficient due to the self-consistent  nature of DSEs.

      In Dyson-Schwinger studies of Landau gauge Yang-Mills propagators in four
dimensions  especially spurious quadratic divergences caused problems, since they
are related to the breaking of gauge symmetry by the regularization/truncation
scheme. Successful treatements were, {\it e.g.}, identification and subtraction in
the relevant tensor-components of the gluon
self-energy \cite{Brown:1988bn,Atkinson:1997tu}, 
and the construction of explicit subtraction terms within the integral kernels
\cite{Fischer:2003rp,Fischer:2003zc,Fischer:2008uz}. 
On the other hand, logarithmic divergences can be treated straightforwardly 
in a MOM-scheme.

      In studies of the Landau gauge gluon propagator DSE truncations  were chosen
such that all terms  containing a four-gluon interaction have been neglected, see,
however, ref.\  \cite{Bloch:2001wz}. A special role is played by the tadpole
diagram:  It only contributes a quadratically divergent constant, which is then
removed in the renormalization process. The other  terms with a four-gluon
interaction are of two-loop order, the so-called sunset and squint diagrams.   As
the gluon-dressing function resulting from DSE studies is in the  intermediate
momentum region somewhat smaller than  the ones obtained from lattice  calculations
or the Functional Renormalization group \cite{Fischer:2008uz}  it has been
speculated that the two-loop terms might provide the missing contribution.
Therefore it is desirable to include the sunset and the squint diagram into DSE
studies of the Yang-Mills propagators propagators in Landau gauge.

      Most non-perturbative studies of the Yang-Mills propagators have been done in
Landau gauge. Another covariant gauge, where in the last years progress has been
gained, is the Maximally Abelian gauge (MAG)
\cite{Shinohara:2003mx,Bornyakov:2003ee,Gracey:2005vu,Capri:2006vv,Huber:2009wh,Alkofer:2011di,Gongyo:2012jb}.
This gauge is especially appealing, since it allows to study the dual
superconductor picture of the Yang-Mills vacuum. In this picture, confinement is
realized via a dual Meissner effect and color-electric fields are squeezed into
flux-tubes by the screening property of the vacuum. This relates to the hypothesis of
Abelian dominance \cite{Ezawa:1982bf} which postulates a dominance of the abelian 
degrees of freedom over the non-abelian ones in the infrared. This hypothesis could 
also be confirmed  in an
infrared analysis of the untruncated Dyson-Schwinger equations in this gauge
\cite{Huber:2009wh}. This investigation showed that the abelian (diagonal) gluons
are enhanced in the infrared in contrast to the off-diagonal gluons and ghosts,
which are infrared suppressed. Hereby, the infrared leading diagram in the gluon
propagator DSE in MAG is the sunset and possibly the squint
diagram. However, both, sunset and squint diagrams, include four-point 
interactions.

      In these proceedings we report on progress which has been gained in
including the sunset diagram into Dyson-Schwinger studies of Yang-Mills theory in
four dimensions. A main step is the identification and subtraction of
overlapping quadratic divergences. We present results on the contribution of
the sunset diagram to the gluon dressing function in Landau gauge and propose a
possible minimal truncation of the propagator DSEs in the MAG.

      \begin{figure}
      \centering
      \includegraphics[width=\textwidth]{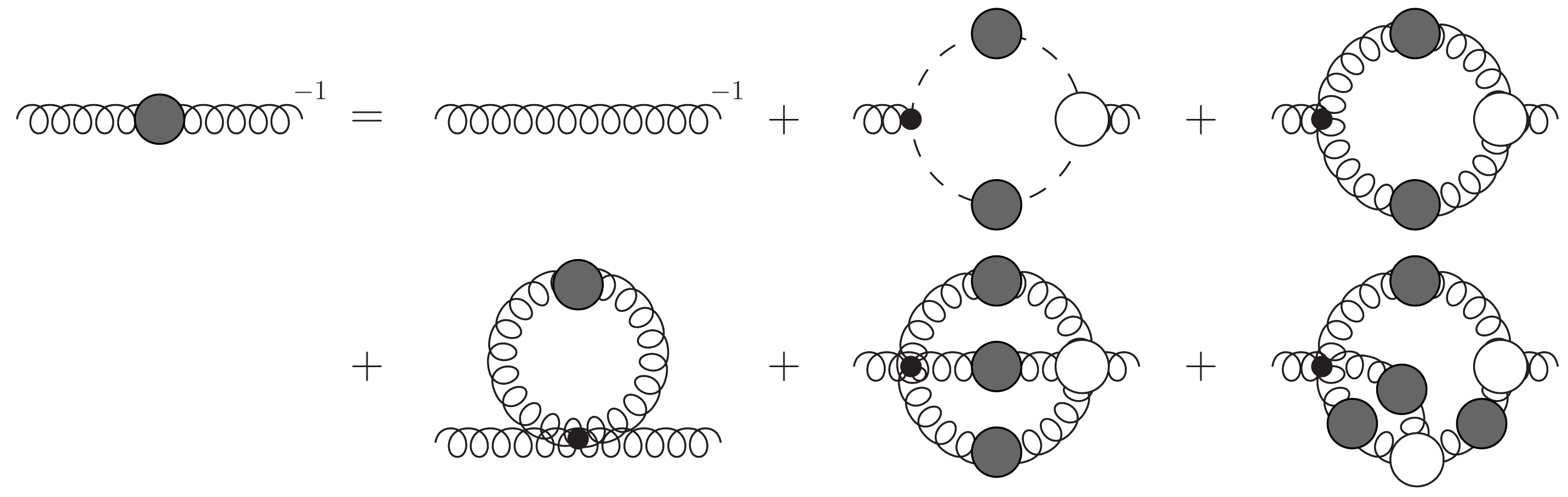}
      \caption{The Dyson-Schwinger equation of the gluon-propagator in Landau gauge.\label{fig_gluonDSE}}
      \end{figure}

  \section{Dyson-Schwinger equations in Landau gauge}
      
      The Dyson-Schwinger equation for the gluon propagator in Landau gauge is
depicted in Fig.\ref{fig_gluonDSE} Over the last decade a truncation scheme has
evolved in which all terms containing four-gluon interactions were excluded. In
addition the ghost-propagator DSE was taken into account , which left only two
unknown Green-functions in the system, the ghost-gluon and three-gluon vertex,
which were modeled.\footnote{See, however, the recent investigation
\cite{Huber:2012kd} in which a dynamic ghost-gluon vertex have been taken into
account.} We want to start from this truncation with the vertex models as introduced
in Ref.\ \cite{Fischer:2003rp} and then add the sunset diagram to this system. 

      The presented truncation, and the usage of a hard momentum cut-off,
introduce quadratic divergences. One possibility to deal with these divergences is
to modify the integration kernel in the gluon-loop such that they cancel the
quadratic divergences from the ghost-loop \cite{Fischer:2003rp}. The corresponding
renormalization condition is the vanishing of the gluon pole mass, 
$Z_m m^2 = 0$, which holds
for any order in perturbation theory\footnote{Note that this is not in
contradiction to the family of decoupling solutions 
\cite{Aguilar:2008xm,Boucaud:2011ug,Fischer:2008uz}
of Landau gauge DSEs which are confirmed by lattice calculations
\cite{Sternbeck:2006rd,Cucchieri:2008fc,Bogolubsky:2009dc,Boucaud:2011ug}.
There one observes infrared-screening, {\it i.e.}, a 
Debye mass which for the scaling solution would just be infinite.}. 
The logarithmic divergencies are then
subtracted in a MOM-scheme. When now including the sunset new overlapping
quadratic divergences appear. Since there is no partner to cancel  these
divergences, we have to subtract the quadratic divergences within the diagram
itself. To fulfill the corresponding renormalization condition we have to
completely subtract these divergences without any additive constant. As it will
be seen below, for a specific momentum partitioning we are able to construct a
modified integration kernel which yield an integral without any quadratic
divergences.

  \section{The sunset diagram: Ultraviolet (UV) analysis}
      
      The analytic expression of the sunset diagram in the DSE of the gluon
propagator is given by
      \begin{equation}
	\Pi_{\mu\nu \,\,\mathrm sunset  }^{ab} (p)
	= - Z_4 \frac 1 6 \int\!\dbar^4q_1\int\!\dbar^4q_2 
	\Gamma^{(0)\,arst}_{\mu\rho\sigma\tau}
	\Gamma^{br's't'}_{\nu\rho'\sigma'\tau'}(-p,p_1,p_2,p_3)
	D^{ss'}_{\sigma\sigma'}(p_1)D^{rr'}_{\rho\rho'}(p_2)
	D^{tt'}_{\tau\tau'}(p_3) 
      \end{equation}
with the bare and fully dressed four-gluon vertices, $\Gamma^{(0)}$
and $\Gamma$, resp., and the fully dressed gluon-propagator $D$. The $p_i$ denote 
the gluon momenta, and the $q_i$ some appropriately chosen loop momenta, see
below. We have also used the short-hand notation 
$\int\!\dbar^4q = \int\!d^4q / (2\pi)^4$.   
To proceed we have to assume a model
for the dressed four-gluon vertex. In this study 
we assume a tree-level tensor structure mulitiplied with a scalar dressing
function which  will be modeled such that the correct UV and IR behavior
is reproduced,
$\Gamma^{abcd}_{\nu\alpha\beta\gamma}(-p,p_1,p_2,p_3)\approx
D_\Gamma(-p,p_1,p_2,p_3) \Gamma^{(0)abcd}_{\nu\alpha\beta\gamma}$. 
Inserting also
the definition for the gluon propagators 
$D^{ab}_{\mu\nu}(q) = \delta^{ab} T_{\mu\nu}(q) {Z(q^2)}/{q^2}$ 
with the transverse projector $T_{\mu\nu}(q) = \delta_{\mu\nu} -
{q_\mu q_\nu}/{q^2}$ and projecting the equation with the generalized
Brown-Pennington projector, $T_{\mu\nu}^\zeta(p) =  \frac 1 3
\left(\delta_{\mu\nu} - \zeta {p_\mu p_\nu}/{p^2}\right)$, and $\delta^{ab}$ 
one obtains
      \begin{equation}
	\Pi_{\mathrm sunset}^{\mathrm proj} = 
	-Z_4 \frac{1}{6} g^4 N_c^2 \int \dbar^4 q_1  
	\int \dbar^4 q_2 \mathcal T D_\Gamma(-p,p_1,p_2,p_3)\,
	\frac{Z(p_1^2)Z(p_2^2)Z(p_3^2)}{p_1^2\, p_2^2 \, p_3^2}
      \end{equation}
      with the tensor $\cal T$ being given by
      \begin{align}
      \mathcal T:=& \,\, 45 + 9 z_{12}z_{13}z_{23} - 
      \zeta \bigl( 9 + 9 \left( z_{01}z_{03}z_{13} + z_{01}z_{02}z_{12} 
      + z_{02}z_{03}z_{23} \right) \nonumber \\
      & \,\, + 3 \left( z_{01}^2 + z_{02}^2 + z_{03}^2 + z_{12}^2 + z_{13}^2 
      + z_{23}^2 - z_{01}^2 z_{23}^2 - z_{02}^2 z_{13}^2 -z_{03}^2 z_{12}^2   
      \right)  \\
      & \,\, + 3 \left( z_{01} z_{02} z_{13} z_{23} + z_{01} z_{03} z_{12} z_{23} 
      + z_{02} z_{03} z_{12} z_{13}  \right) \bigr) \,,\nonumber
      \end{align}
      and $z_{ij} := p_i\cdot p_j / \sqrt{p_i^2p_j^2}$. 
      Choosing now a specific momentum partitioning, 
      $p_1 = q_1, p_2 = p+q_2, p_3 = -q_1-q_2$, and integrating out those 
      angles which do not appear in the integrand we arrive at
      \begin{multline}
	\Pi_{\mathrm sunset}^{\mathrm proj} = - \frac{Z_4g^4 N_c^2}{3(2\pi)^6} 
	 \int_0^\Lambda dq_1 \int_0^\Lambda dq_2 \int_{-1}^1 dz_1 
	 \sqrt{1-z_1^2} \int_{-1}^1 dz_2 \sqrt{1-z_2^2} \int_{-1}^1 dy \\
	 q_1q_2^3\, \mathcal T \, D_\Gamma(-p,q_1,p+q_2,-q_1-q_2)\,\,
	 \frac{Z(q_1^2)Z((p+q_2)^2)Z((q_1+q_2)^2)}{ (p+q_2)^2 (q_1+q_2)^2} 
	 \label{sunset} \,.
      \end{multline}

      To gain insight into the UV behavior of \eqref{sunset} a method which has
proven to be very useful in the UV analysis of DSEs in asymptotically free
theories is the so-called y-max approximation
\cite{Atkinson:1997tu,Fischer:2003rp}. In the high-momentum regime the
gluon dressing function behaves according to the perturbative logarithmic
behavior. Due to this weak momentum dependence one can then safely  neglect the
dependence of the dressing functions on the angles.  Since the cut-off can be
choosen arbitrarily large the largest contributions  to the UV behaviour will
originate from the momentum region $\Lambda>q_1,q_2 > p$. For the sunset 
additionally we have to assume that exceptional momenta will not contribute
significantly to the integral which for the employed non-perturbative gluon
propagators is justified due to their infrared (IR) suppression. Here we implement
a scaling solution, which even vanishes for $p^2\rightarrow 0$. Under these
assumptions one can integrate over the angles analytically and obtains
      \begin{equation}
	\Pi_{\mathrm sunset}^{\mathrm ymax} =  
	- \frac{Z_4g^4 N_c^2}{3(2\pi)^6}  \int_p^\Lambda dq_1 
	\int_p^\Lambda dq_2 \\D_\Gamma(-p,q_1,p+q_2,-q_1-q_2)\,
	Z(q_1^2)Z((p+q_2)^2)Z((q_1+q_2)^2) \,I_{\gtrless} \,.
      \end{equation}
	For $q_1>q_2$ the integrand $I_{\gtrless}$ is given by,
      \begin{subequations}\label{Ismallbig}
      \begin{equation}
	I_{\gtrless} = - \frac{189\pi^2}{32} \frac{q_2}{q_1} 
	(\zeta - 4) +\frac{9\pi^2}{32} \frac{q_2^3}{q_1^3} 
	(\zeta - 4) - \frac{9\pi^2}{32} \frac{p^2q_2}{q_1^3} (\zeta - 1)
      \end{equation}
      and for $q_2>q_1$ one gets,
      \begin{multline}
	I_{\gtrless} = - \frac{189\pi^2}{32} \frac{q_1}{q_2} 
	(\zeta - 4) +\frac{9\pi^2}{32} \frac{q_1^3}{q_2^3} (\zeta - 4) - 
	\frac{27\pi^2}{32} \frac{p^2q_1}{q_2^3} \\ + \frac{3\pi^2}{16} 
	\frac{p^2q_1^3}{q_2^5} (6+\zeta) - \zeta \frac{15\pi^2}{32} 
	\left( \frac{p^4q_1^3}{q_2^7} + \frac{p^2q_1^5}{q_2^7} - 
	\frac{p^4q_1^5}{q_2^9}\right)\,.
      \end{multline}
      \end{subequations}
      
      The explicit expressions \eqsref{Ismallbig} allow to identify the 
quadratic divergences
and construct a subtraction term such that upon integration the quadratic
divergences are canceled for both cases. This procedure provides a 
``regulated'' tensor structure,
      \begin{equation}
	 \widetilde{\mathcal T} = \mathcal T - 
	 \left( \zeta - 4\right)\left(-\frac{225}{16} + \frac{9}{4} (z_{12} + 
	 3 z_{01}) \right)\,. \label{subtraction}
      \end{equation}
      If we replace $\mathcal T$ in \eqref{sunset} by $\widetilde{\mathcal T}$, 
      no quadratic divergences will appear.

 A few comments are in order here. First note that the quadratic divergences come
with a factor $\zeta - 4$. This means that the quadratic divergences of the sunset
only contribute to the $\delta_{\mu\nu}$-component of the gluon polarization
tensor as is expected from general considerations \cite{Brown:1988bn}.  Second, in
the past two-loop diagrams were not included due to the complications  provided by
their overlapping divergences.  However, using the regulated  tensor structure
\eqref{subtraction} the quadratic divergences are cancelled directly.
As the explicit form of this subtraction term depends on the particularly
chosen momentum partitioning this has become possible due to the specific choice
taken, and whether a similar subtraction is possible for general
momentum partitionings remains to be clarified.

  \section{A model for the four-gluon vertex dressing}
      
      The unknown required input into \eqref{sunset} is the dressing function of
the four-gluon vertex. While the UV behavior of this function is known from
perturbation theory only little is known when one of the momenta becomes
small.\footnote{To our knowledge there is no lattice study of this object.}  In
refs.\ \cite{Fischer:2006vf,Kellermann:2008iw} an infrared behavior of the
four-gluon vertex consistent with the scaling solutions of the DSEs has been 
determined. Under consideration of the available information we model the
vertex dressing function as a product of the propagator dressing functions:
      \begin{equation}
	 D_\Gamma(-p,p_1,p_2,p_3) = 
	 \frac{1}{Z_4} \frac{\left[G(p_1^2)G(p_2^2)\right]^\alpha}
	 {Z(p_3^2)\,\left[Z(p_1^2)Z(p_2^2)\right]^{1-\beta}} \,,
      \end{equation}
where we introduced two parameters $\alpha$ and $\beta$ which are 
determined by the UV and IR behaviour. For high energies the dressing function 
are required to obey a logarythmic scaling 
$D_\Gamma \sim (\log x)^{-\gamma+2\delta}$ 
with the anomalous dimensions of the gluon propagator, $\gamma = -\frac {13}{22}$, 
and the ghost propagator $\delta = -\frac{9}{44}$, respectively. 
To be consistent with the scaling solution in the IR we  require
$D_\Gamma \sim (x)^{-4\kappa}$ with the IR scaling exponent 
$\kappa \approx 0.59535$ \cite{Lerche:2002ep,Fischer:2006vf}. 
Thus we can determine the parameters to 
$\alpha = -4\delta$ and $\beta = \frac 1 2 (1-4\delta)$.
      
 In this first preliminary investigation this model has been chosen mainly due to
technical reasons. Besides the missing Bose symmetry  the different treatment of
the propagators even within the sunset diagram calls for improvement.  As we
divide out the dressing function  with argument $p_3$ this propagator becomes
undressed which allows us to integrate over two of the three angles  analytically.
This saves a substantial amount of  computing time. We will improve on this
shortcoming via transporting these numerical calculations onto GPUs along the
lines of Ref.\ \cite{Hopfer:2012ht}. This should enable us to 
calculate all angular integrals numerically which is a prerequisite 
to keep Bose symmetry in the employed four-gluon-vertex model.

      For the numerical calculation one still needs to choose a partitioning 
of the two radial integrals. In addition, there exist integrable divergences for 
$q_1- q_2 \to 0$ which have to be treated seperately. Details will  be published 
elsewhere.

  \section{Results and outlook}
      
      \begin{figure}
      \centering
      \includegraphics[width=.48\textwidth,bb=0 0 792 612]{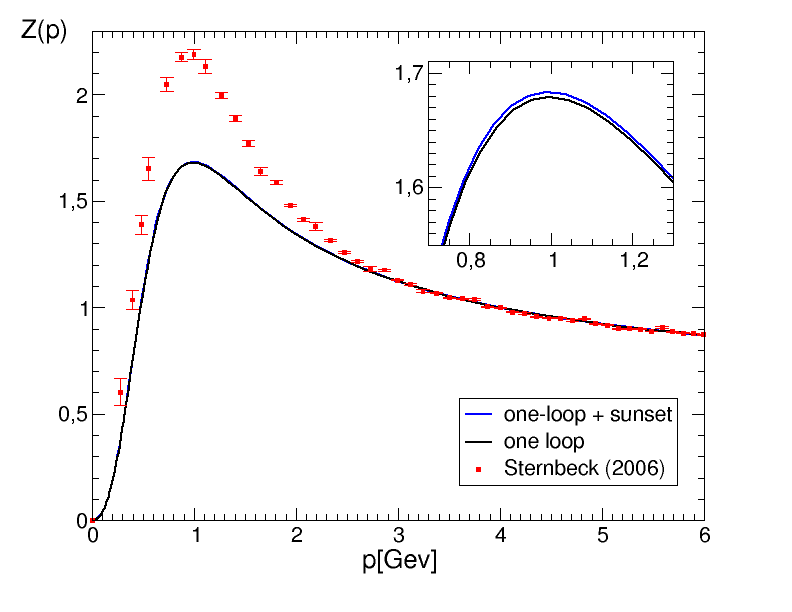}
      \includegraphics[width=0.48\textwidth,bb=0 0 792 612]{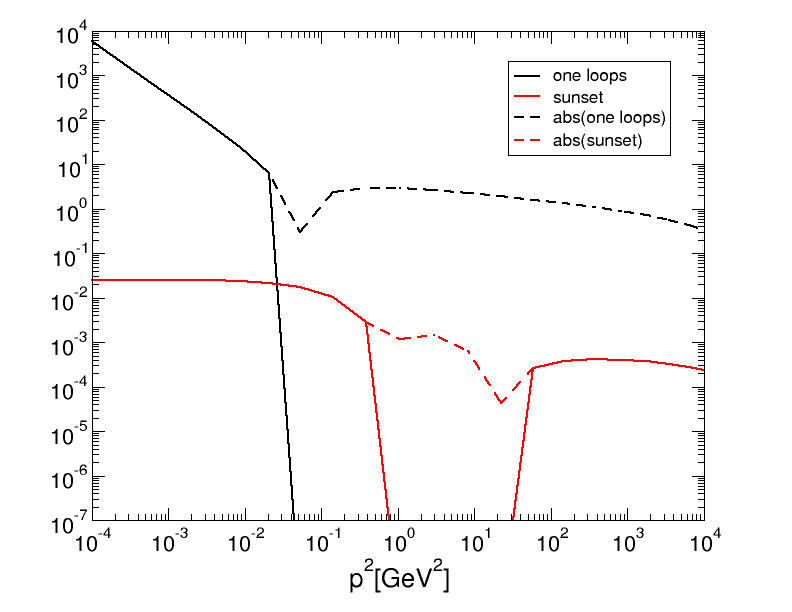}


      \caption{\label{fig_latcomp} \emph{Left: }The Landau gauge 
      gluon dressing function with (blue) and without (black) contributions
      from the sunset diagram compared to lattice data of Ref.\ 
      \cite{Sternbeck:2006rd}. 
      \emph{ Right: } Contributions of the one-loop diagrams and the sunset 
      after subtracting quadratic divergencies.}
      \end{figure}
      
Within the approximations described above we included the sunset diagram into a
Landau gauge DSE study. As one can see from Fig.\ref{fig_latcomp} the influence of
the sunset diagram is astonishingly small. This is in contrast to  expectations
that the two-loop terms should contribute significantly in the mid-momentum
regime. Even our preliminary results provide compelling evidence  that the sunset
diagram is completely unimportant in the gluon propagator DSE.  At this point two
remarks are in order: First, the precise form of the three-guon vertex in the
one-gluon-loop term is very important as can be seen from very recently obtained
results  \cite{Huber:2012kd} also presented on this conference.\footnote{See,
however,  also Refs.\ \cite{Maas:2004se} for a corresponding discussion of the
importance of the one-gluon-loop term.} An improvement of the
three-gluon vertex model, especially the inclusion of a sign flip at small
momenta, leads to an enhancement of the gluon dressing function in the
mid-momentum regime. 
Second, preliminary results on including 
the squint diagram in the gluon propagator DSE \cite{HMA13} 
point towards a small but still sizeable contribution.
Here the use of a Bose symmetric four-gluon dressing
function is then mandatory to allow for definite conclusions.

  \section{Implications on the Maximally Abelian gauge}

	\begin{figure}
	\centering
	\includegraphics[width=\textwidth]{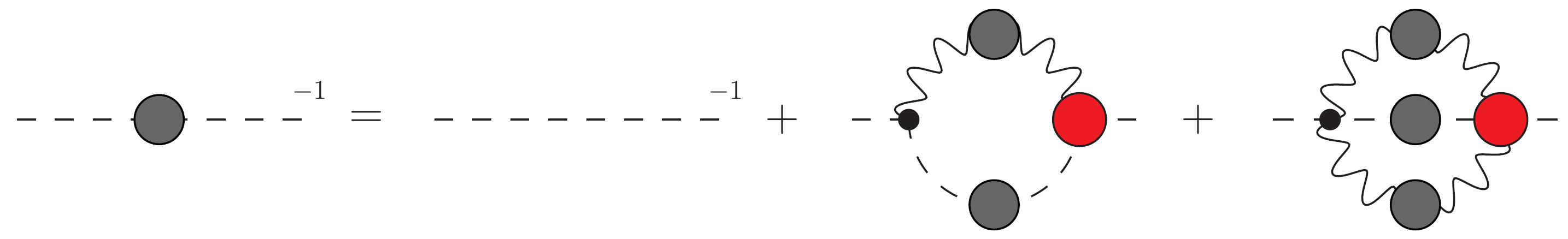}\hspace{4mm}
	\includegraphics[width=\textwidth]{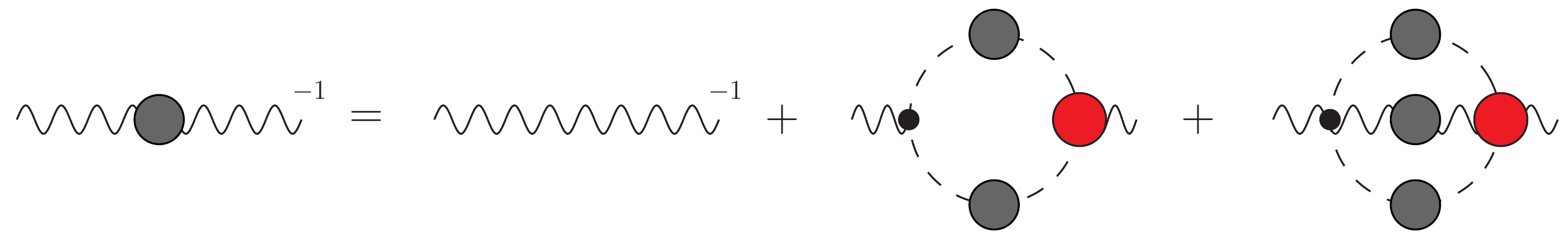}
	\caption{The proposed equations for ghosts and diagonal gluon 
	in the MAG.\label{fig:MAG}}
	\end{figure}

As mentioned an IR analysis of the Yang-Mills propagator DSEs and Renormalization
Group Equations in the Maximally Abelian gauge showed \cite{Huber:2009wh}  that
the sunset diagram is an IR leading term. We propose here a minimal set of coupled
equations for the propagators of ghost and diagonal gluon in the Maximally Abelian
gauge which is depicted in Fig.~\ref{fig:MAG}.
It is the minimal truncation of the corresponding DSEs which contains the
leading UV and IR behaviour. The corresponding one-loop terms give the correct UV
behavior \cite{Gracey:2005vu}, where as the sunset diagrams give the IR behavior
\cite{Huber:2009wh}. Also there are only two vertices which have to be modeled to
close the system, the three-point $(A\bar c c)$- and the four point $(AA \bar c
c)$-vertex. Again we assume a tree-level structure for these vertices dressed
with a scalar function which reproduces the correct UV and IR behaviour of these
Green functions. The quadratic divergences can be subtracted along the same line
as presented above: A y-max approximation allows for the construction of a
regularized tensor structure in the integral kernels. The logarthmic divergences 
are then treated in a MOM-scheme.
 Numerical studies of this set of equations as a starting point of an
investigation of the Yang-Mills propagators in the Maximally Abelian gauge are
in progress.

  \section{Acknowledgements}

We are grateful to the organizers of the {\it Xth Quark Confinement and the Hadron
Spectrum} conference for all their efforts which made this extraordinary event
possible.\\
      We thank A.~Sternbeck for providing us lattice results. VM is
funded by the Austrian Science Fund, FWF, through the Doctoral Program on Hadrons
in Vacuum, Nuclei, and Stars (FWF DK W1203-N16).


\end{document}